\documentclass[conference,compsoc]{IEEEtran}

\usepackage[utf8]{inputenc} 
\usepackage[T1]{fontenc}    
\usepackage{hyperref}       
\usepackage{url}            
\usepackage{booktabs}       
\usepackage{amsfonts}       
\usepackage{nicefrac}       
\usepackage{microtype}      
\usepackage{xcolor}         
\usepackage{wrapfig}
\renewcommand{\paragraph}[1]{\bigskip\noindent\textbf{#1}}
\usepackage{caption}
\usepackage{subcaption}
\usepackage{threeparttable} 

\usepackage[utf8]{inputenc}
\usepackage{amsmath}
\usepackage{amsfonts}
\usepackage{graphicx}

\newcommand{\citeflashattention}{\cite{dao2022flashattention}}
\newcommand{\citespeculative}{\cite{stern2018blockwise}}
\newcommand{\citespecinfer}{\cite{miao2024specinfer}}
\newcommand{\citespeculativestreaming}{\cite{bhendawade2024speculative}}

\newcommand{\citemedusa}{\cite{cai2024medusa}}
\newcommand{\citeparallel}{\cite{santilli2023accelerating}}

\newcommand{\citeconsistency}{\cite{kou2024cllms}}

\title{Remote Timing Attacks on Efficient Language Model Inference}

\author{%
Nicholas Carlini \qquad Milad Nasr \\
\emph{Google DeepMind}
}

\begin{document}

\maketitle

\begin{abstract}
    Scaling up language models has significantly increased their capabilities.
    But larger models are slower models, and so there is now an extensive body
    of work (e.g., speculative sampling or parallel decoding)
    that improves the (average case) efficiency of
    language model generation.
    But these techniques introduce data-dependent timing characteristics.
    We show it is possible to exploit these timing differences to 
    mount a \emph{timing attack}.
    By monitoring the (encrypted)
    network traffic between a victim user and a remote language model, we
    can learn information about the content of messages by noting when responses
    are faster or slower.
    With complete black-box access,
    on open source systems we show how it is possible to learn the topic of a user's conversation (e.g., medical advice vs. coding assistance) with 90\%+ precision,
    and on production systems like OpenAI's ChatGPT and Anthropic's Claude we can
    distinguish between specific messages or infer the user's language.
    We further show that an active adversary can leverage a boosting
    attack to recover PII placed in messages
    (e.g., phone numbers or credit card numbers) for open source systems.
    We conclude with potential defenses and directions for future work.
\end{abstract}

\section{Introduction}

In order to support increasingly advanced capabilities, 
language models are rapidly increasing in size.
Just five years ago the largest language models had one billion parameters---today,
the largest models contain over one trillion parameters.
This scale necessitates that large models be deployed in specialized 
data-centers on hardware designed for performing language model inference;
and so to access these models users must make queries of these remote systems
and can not easily deploy them on their own machines.

This increase in model size also increases both the monetary cost of serving models
and also increases the inference latency.
But \emph{expensive} and \emph{slow} are undesirable properties of products.
And so there has been significant work directed towards 
developing techniques to efficiently serve language models without degrading utility.

Classical techniques improve performance uniformly across all queries.
Low-precision inference evaluates the expensive matrix multiplication
over reduced-precision domains, e.g., float-16 \cite{google2019bfloat16} or float-8 \cite{nvidia2023fp8}.
Techniques such as weight quantization \cite{dettmers2022llmint8} reduce the memory requirements
of storing the model in RAM and thereby allows for larger batches of
examples, again improving speed and reducing cost.
And methods like FlashAttention \cite{dao2022flashattention} develop hardware-aware 
methods to improve efficiency by accessing data in a memory efficient manner.

Recent techniques take a different approach; and
instead of directly improving the underlying numerical efficiency of the
computation that a model performs,
develop methods that improve the algorithms themselves.
For example, recent work has demonstrated that it is possible to perform
an adaptive amount of compute per token, depending on the difficulty of that
token.
It is easier, for instance, to predict the token following
``1 + 2 = '' than for ``log(53.5) = ''.
Other work draws inspiration from speculative execution in processors, and 
predicts future tokens many steps into the future, and then
checks if these predictions were correct (and then repeats the process),
or ``rolls back'' and takes only the next one token as a standard model would do if the prediction was wrong.
Importantly, each of these techniques introduces data-dependent timing
characteristics into the inference procedure:
it is thus possible for models to behave \emph{slower} on one query
because it is hard, and \emph{faster} on another because it is easy.

\paragraph{Contributions.} 
We introduce the first timing attacks on efficient language model inference.
Under strong assumptions, we show how a network adversary can recover
PII contained in the model's prompt.
And under practical assumptions, we show how a passive network adversary can
learn limited information (e.g., the topic of conversation) about the conversation of users querying OpenAI's GPT or Anthropic's Claude family of language models.

Classical timing attacks on systems are remarkably difficult to exploit given
the slight (e.g., microsecond) difference in time a computation takes to perform, e.g., depending on whether or not a specific memory address is in the in L1 cache.
But because language models are much slower (and often take fractions of
a second to generate a single token),
our attacks can easily be exploited over the network.

Specifically, we make the following contributions:
\begin{itemize}
    \item We formalize the threat model of timing attacks on efficient language model inference.
    \item We demonstrate that production language models today use efficient inference techniques,
    and demonstrate they are vulnerable to our attacks.
    \item We construct the first timing attacks in both the passive and active settings.
    \item We propose and evaluate several defenses that can completely prevent our attacks.
\end{itemize}

\paragraph{Responsible Disclosure.} We disclosed our attack to OpenAI in January 2024, and Anthropic in April 2024.

\begin{table*}[htbp]
\centering
\begin{threeparttable}
\caption{Comparison of Efficient Language Model Inference Methods.}
\label{tab:methods}
\label{tab:method_comparison}
\begin{tabular}{@{}llcp{9cm}@{}}
\toprule
\textbf{Method Name} & \textbf{Data-Dependent} & \textbf{Vulnerable?} & \textbf{Key Idea} \\ 
\midrule
Distillation~\cite{hinton2015distilling} & No & No & Train an efficient small model to mimic a larger model. \\
Quantization~\cite{jacob2018quantization} & No & No & Use low precision arithmetic, saving memory and using faster hardware. \\
Flash Attention~\cite{dao2022flashattention} & No & No & Compute attention in a hardware-aware manner to maximize cache efficiency.\\
\midrule
Speculative Decoding~\cite{stern2018blockwise} & Yes & Yes$^1$ & A small \emph{draft} model predicts multiple tokens at a time, and the large \emph{target} model (in a single evaluation) accepts or rejects the predicted tokens. \\
SpecInfer~\cite{miao2024specinfer} & Yes & Yes$^1$ & Like speculative decoding, but trains a boosted collection of small speculative models and performs a tree-based verification. \\
Speculative Streaming~\cite{bhendawade2024speculative} &  Yes & Yes$^2$ & Like speculative decoding, but trains a language model to predict multiple tokens at a time and uses the future predicted tokens as speculative. \\
Mixture of Depths~\cite{raposo2024mixture} & Yes & Yes$^2$ & Trains a model to conditionally perform more or less computation by skipping certain layers of the transformer when it is deemed unnecessary. \\
Medusa~\cite{cai2024medusa} & Yes & Yes$^1$ & Like speculative streaming, but trains multiple independent prediction heads and so is more accurate but has higher memory overhead.  \\
Parallel Decoding~\cite{santilli2023accelerating} & Yes & Yes$^2$ & Uses Jacobi-iteration to find a fixed-point solution to predicting future tokens by repeatedly querying a single model on multiple future tokens. \\
Lookahead Decoding~\cite{fu2024break} & Yes & Yes$^1$ & An improvement of parallel decoding using historical k-gram predictions to more rapidly converge on the Jacobi-iteration fixed-point. \\
Consistency language models~\cite{kou2024cllms} & Yes & Yes$^1$ & A further improvement of parallel decoding with Jacobi-iteration by fine-tuning the model to more accurately predict blocks of tokens at a time. \\

\bottomrule
\end{tabular}
\begin{tablenotes}
\item[1] The official implementation is public, and we verify it is vulnerable to timing attacks in \S\ref{sec:attack5}.
\item[2] The implementation is not public, but by studying the algorithm we are confident it will be vulnerable.
\end{tablenotes}
\end{threeparttable}

\vspace{1em}
\end{table*}

\newpage
\section{Background}

\paragraph{Large Language Models} (LLMs) are the basis of most modern NLP applications \cite{vaswani2017attention,devlin2018bert}.
While the exact techniques used have varied over time,
one key fact of language models has remained constant:
they generate text one ``token'' (e.g., a word or sub-word) at a time by
modeling the data distribution $Pr(x_i | x_1,\dots,x_{i-1})$,
and then \emph{autoregressively} sampling future tokens.

Language models have grown rapidly in size over the past several years---from
1 billion parameters in 2019 to 1 trillion parameters in 2023.

For the purpose of this paper it is necessary to understand two properties of
current \emph{transformer-based} language models.
\begin{itemize}
  \item \textbf{Generation is sequential.}
This autoregressive sampling has a significant drawback: text generation
is inherently \emph{sequential}. 
The $i$th token must be generated before
the $i+1$st token can begin being processed.

    \item \textbf{Scoring is parallelizable.}
    Due to the design of the transformer architecture, the cost to compute
    $Pr(x_i | x_1,\dots,x_{i-1})$ is almost identical to the cost
    to compute $Pr(x_j | x_1,\dots,x_{j-1})$ for all $j \le i$.
    Therefore, \emph{verifying} that a sequence of tokens has been correctly
    generated is much faster than generating that same sequence.
\end{itemize}

\paragraph{Efficient LLM inference.}
Language models employ a number of techniques to reduce their response latency.
As mentioned previously, traditional efficiency improvement techniques like weight/activation
quantization or methods like FlashAttention~\citeflashattention{} that are designed to utilize the architecture of hardware used in running the LLMs. 

But language models make new efficient inference techniques possible. 
One of the simplest and most widely adopted approaches to minimizing latency is to have the 
language model \emph{stream} its response
to the user token-by-token.
This allows the user to start to read a (partial) reply before the complete output is available.
This is the approach taken by all of the largest language model providers,
including OpenAI, Google, and Anthropic.

Other techniques, like
the memory-layout-aware Flash Attention~\cite{dao2022flashattention} have provided
significant gains.
But developing methods that improve the efficiency of models for all inputs uniformly
has become increasingly difficult.

Recent work has thus turned to developing methods that improve LLM efficiency \emph{on average},
potentially at the cost of increasing (or, not reducing) latency in the worst case.
As mentioned earlier, these methods rely on the observation that some predictions are ``easier'' 
and can therefore be performed with less work than others.
But this introduces a timing side channel, which we show is exploitable in this paper.

Table~\ref{tab:method_comparison} lists several recent advances in
efficient language modeling, along with whether or not the technique introduces data-dependent
timing characteristics.
Of the eight methods we identify here, five have public implementations we are able to
reproduce, and which we demonstrate are all vulnerable to the timing attacks we develop.

A complete description of each of these methods is not possible due to the space
constraints of this paper.
Fortunately, for the purpose of our attacks in the remainder of this paper,
it will not be necessary to understand
exactly how each of these methods work.

\paragraph{Example: speculative decoding.}
Nevertheless, to provide intuition for how efficient inference techniques
introduce data-dependent properties, we provide a complete description of 
\emph{speculative decoding} (also known as \emph{speculative sampling}) here.
Consider a model provider who has two models: a large, powerful, and slow model
$f_{big}$ and a small and efficient model $f_{small}$.
Whenever we would like to query the large model on the prompt $p$,
we first query the small model on the same prompt.
Once we have generated $k$ tokens with the small model, we then
invoke the large model \emph{once} to check the work of the small model
on all $k$ tokens simultaneously.
Importantly, the processing time necessary to evaluate $k$ tokens is (much)
faster than the time needed to sequentially generate $k$ tokens.

Then, drawing inspiration from speculative execution,
if the large model agrees with the small model on these new tokens, we \emph{accept}
this token sequence and repeat the procedure.
However, if there is some token at which the large model and small model disagree,
we \emph{roll back} to the point of disagreement (up to and including the very first
token) and repeat from this location.
Empirically, speculative decoding can double the speed of large language models.

\paragraph{Traffic Analysis Attacks}
Traffic analysis focuses on extracting sensitive information from the characteristics of network communication, like packet sizes, timings, and their derivatives. 
Traffic analysis is particularly important when encryption and other content-masking techniques prevent direct inspection of the communication content. 
The common types of the traffic analysis techniques are protocol/website fingerprinting~\cite{crotti2007traffic} where they are mainly used to extract the protocol a victim is using or the website a victim is visiting, when the victim is using some encrypted proxy system (such as VPNs or anonymous networks). Alternatively, in \emph{flow correlation attacks} the goal is to correlate traffic going to and coming out of a ``mixnet''. 
Other than common traffic analysis applications, traffic analysis attacks have been applied to other application scenarios in order to
extract sensitive information.
Various types of side channel attacks are built on traffic analysis.
For instance, multiple studies
use traffic analysis to uncover not only the language spoken over encrypted VoIP connections~\cite{wright2007language},
but also the spoken phrases~\cite{wright2008spot,white2011phonotactic}.
Chen et al.\cite{schuster2017beauty} use traffic analysis to infer the video streams being watched over encrypted channels. 

Concurrent work \cite{weiss2024your} discovered a simple length side channel
present in several language model services.
Our attack is independent, and remains effective even after
all services patched this paper's length side channel attack.

\section{Threat Model}

We assume an adversary who aims to exploit the timing characteristics
of a language model to learn information that should not otherwise be
visible.
For example, an adversary may wish to learn approximate information,
such as the \emph{language} the user is conversing
with the model (e.g., English or French),
or \emph{the topic of conversation} (e.g., coding assistance or medical advice).
Alternatively, the adversary may wish to learn an exact piece of information
e.g., whether or not the conversation contains the phrase ``my pin is 1234''.

We consider two possible situations that may arise:

\paragraph{Passive Attacks}
In the simplest setting, a victim user is interacting with a language model,
and a network adversary monitors the (encrypted) traffic between the user
and the language model.
This is a common threat model in traffic analysis works~\cite{wright2007language,nasr2017compressive,nasr2018deepcorr,houmansadr2011swirl,mcpherson2016covertcast,wang2014effective,k-fing,juarez2014critical,juarez2016toward}. A network adversary could obtain this capability through many means.
For example, they could be a malicious internet service provider;
could have compromised a router along the path between the user and language model; or
could have a device on the same wireless network as the user.
In each of these cases the adversary can monitor the connection---but not modify any data.

\paragraph{Active Attacks}
In the other setting we consider, the
adversary plays a more active role in the attack.
Specifically, we assume an adversary who can exercise (some) control
over the content of the message sent to the language model---but
still can not view anything about the response other than the time it takes.
(This threat model is similar to that of a \emph{chosen plaintext} attack
on a cryptographic system, where the adversary is presumed to be able to
request the encryption of arbitrary inputs.)
While this threat model may seem to be extremely strong,
there are several scenarios where this threat model is meaningful.

Consider the setting where a malicious company provides a service that (behind the scenes)
makes use of another company's language model.
This service could be anything, for example it could be a research tool designed to help users explore various topics. The malicious company controls the exact template used to call the language model's API. While the user's input remains private on their device, the malicious company can still monitor network metadata, like the time it takes for a response. The target end user's device can send multiple requests to the model and manipulate parts of those requests, but they cannot see the actual responses. Think of it like a "blind SQL injection" attack, where the adversary tries to extract sensitive information by crafting a sequence of queries to the remote system, relying on subtle variations in the responses to deduce the hidden data. Here, the malicious company, much like a traditional adversary attempting SQL injection, sends a series of crafted prompts through their seemingly innocuous research tool, probing the language model for vulnerabilities and attempting to glean confidential information based on the timing and patterns of the responses they receive. While research topics might not be very sensitive, this attack can be applied to extract more sensitive information.

\paragraph{Formalism.}
We assume a user $\mathcal{U}$ and interacts with a language model
oracle $\mathcal{O}$.
Consider a predicate $\phi$ that can be applied to a conversation;
e.g., $\phi$ may represent the specific question ``is the message exactly [message]'',
something more broad ``is this a about medical advice?'',
or something in between ``does this message contain the number 527?''
The user then flips a coin and decides to either send a message to the model
that satisfies predicate $p$, or with equal probability, does not satisfy $p$.
The model responds and streams the response back to the user as a sequence of tokens $m_i$.
Each of the messages $m_i$ exchanged by the user and model are annotated with
the time $t_i$, and when sent over the network are encrypted by a perfect encryption
algorithm that leaks no information about the content of the message.
The objective of the adversary is to study $\{t_i\}_{1}^{n}$ and
then predict with probability greater than $1\over2$ whether or not
$p(\{m_i\}_1^n)$.

\paragraph{Knowledge.}
Throughout this paper, we consider an adversary who has black-box access
to the system the victim will eventually use.
That is, the adversary is allowed to make an arbitrary (but cost-limited)
number of queries of the system, but assume the adversary does not have
complete unrestricted access to the language models to, e.g., perform
gradient-based attacks.
(In Section~\ref{sec:whitebox} we show that, even if an adversary could get white-box
access to the target language model, current attacks are insufficiently
powerful to make use of these additional capabilities.)

We additionally consider either an entirely passive adversary who can only
view the network traffic between a victim user and a remote language model,
or an active adversary who can control (part of) a user's queries.

\section{Passive Attacks on Open Source Models}
When an adversary sits on the network path, or can otherwise
monitor communication between a victim user and the remote language model,
we can learn information about the encrypted messages.

\subsection{Attack Primitive: A/B Testing}

We begin with our simple attack primitive that performs a hypothesis test
to reliably distinguish between a user sending one of two prompts $p_A$ and $p_B$ to a model.
(Formally: we let $\phi(x) = (x == p_A)$.)
Depending on the exact set of prompts $p_A$ and $p_B$ there may or may not be
any immediate security relevance to this question \emph{per se};
nevertheless, demonstrating that it is possible to learn \emph{anything}
about a user's conversation with a language model by observing the timing
characteristics of the model is a privacy violation.
Our attack has two phases: initialization and inference.

In many ways, this attack is completely analogous to the methods that are
used by network adversaries in the website fingerprinting literature~\cite{juarez2014critical,wang2013improved,rimmer2017automated,sirinam2018deep,cherubin2022online,cai2012touching,wang2020high,panchenko2016website,hayes2016k}. 
Website fingerprinting attacks aim to identify the specific website a user is visiting  when they are using encrypted channels like VPNs or the Tor network. 
This comparison is particularly relevant in the context of closed-world website fingerprinting, where the goal is to differentiate among a limited set of websites, mirroring our focus on two specific prompts. (Later we will generalize this to 1-of-$N$ prompts.)

\paragraph{Phase 1: Initialization}
To begin, we construct a \emph{signature} of the timing of response tokens when querying the
model either with prompt $p_A$ or prompt $p_B$.%
\footnote{To ensure that our attack is relying only on timing characteristics and not, e.g., the length of the message
we request just 150 tokens of output from the model.}
Specifically, we query the the remote model many (e.g., 100) 
times with each prompt $p_A$ and $p_B$, 
and record the inter-token response time.

We then train a classifier to distinguish between these two distributions.
This classifier can be as simple as a single average-and-threshold (if one
query is systematically faster at responding than the other),
or in other cases a Gaussian Mixture Model~\cite{mclachlan1988mixture} or neural network.
Regardless of the exact technique, the classifier $f$ is designed so that
negative values indicate the message was probably $p_A$ and positive values the message was $p_B$;
this allows us to sweep a threshold $\tau$ to compute a precision-recall
plot.

\paragraph{Phase 2: Inference}
At attack time, a victim user makes a request to the language model sending either
prompt $p_A$ or $p_b$.
The adversary then monitors the network traffic and records the inter-packet response time $\{t_i\}$.
From here, the adversary computes the goodness of fit between this measured
distribution $\{t_i\}$ and each of the observed distributions from initialization
using the pretrained classifier.

\subsection{Evaluation Methodology}

We place a language model server on a machine running FastChat\footnote{%
https://github.com/lm-sys/FastChat}, the most popular service that
exposes a language model remotely via a standard API.
We integrate each of the five efficient inference methods we identify that
contain public implementations from Table~\ref{tab:method_comparison}:
Speculative Decoding~\citespeculative,
SpecInfer~\citespecinfer,
Speculative Streaming~\citespeculativestreaming,
Medusa~\citemedusa,
Parallel Decoding~\citeparallel,
and Consistency Large Language Models~\citeconsistency.
Because FastChat exposes an HTTP interface, 
we implement an HTTPS proxy with Nginx so that all communication with the
language model can occur over an encrypted channel to best simulate realistic settings.

We place the victim user inside of a virtual machine on a separate
cloud machine (in our setup, a Docker container),
with the adversary acting as a
 network adversary that can have access to the communication link and can observe the communication between the users and the serving servers.
In all cases, the API sends requests over the HTTPS channel,
which, following the guarantees of a secure channel,
should prevent the adversary from learning anything about the contents of these messages.

The adversary monitors all network traffic sent to or from
the victim's virtual machine;
by inspecting the target IP address and port, it is possible to limit the
packets to just those intended for a specific language model provider.
When the victim makes the language model request, the adversary captures all packets,
discards their (unreadable, encrypted) contents,
records the time at which the packet was received,
and then computes the inter packet delay.

For this attack we train simple Gaussian Mixture Models (GMMs)
on 100 queries in advance of the attack,
but following the same protocol.
Then, at attack time, the adversary computes the goodness of fit
between the observed inter packet delay and each of the two
training distributions.

\subsection{Results}

We evaluate our attacks in two different settings, for each of the five
efficient inference methods discussed above.

Because several of these efficient methods require a model specially-adapted to that technique, it is not meaningful
to draw direct comparisons between different methods---because the
underlying language models may be different.
Instead, our hope in performing these experiments is to demonstrate
that regardless of the exact technique used, all exhibit
\emph{exploitable} data-dependent timing characteristics.

\subsubsection{A/B testing experiments}

We begin with a simple experiment to evaluate the ability of our attack
to distinguish between one of two possible input messages.
Specifically, we send either the message ``generate a sequence of 50 random numbers''
or ``generate the first 50 numbers in order'' and record the packet timing
for each of these queries.

\begin{figure}
    \centering
    \includegraphics[scale=.75]{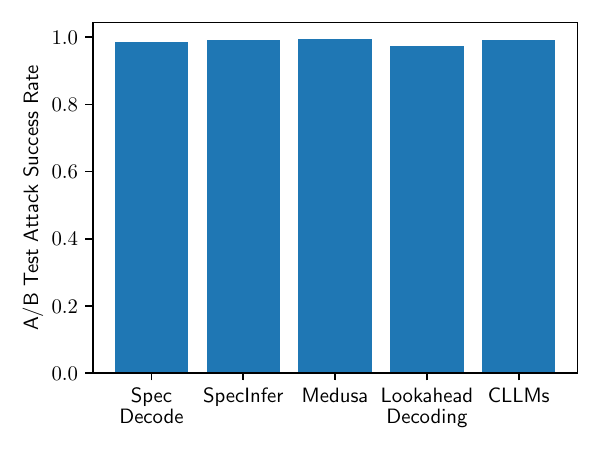}
    \caption{All efficient inference methods we tested are vulnerable to timing attacks,
    and we can reliably distinguish between two queries with near 100\% attack success rate.}
    \label{fig:seq}
\end{figure}

Figure~\ref{fig:seq} shows the result of this experiment.
Overall, we find that the attack success rate is exceptionally high for all
efficient inference methods.
This preliminary experiment confirms that, if an adversary were to have
specific \emph{a priori} knowledge that a victim was going to send one of a small
set of messages, it would be possible to precisely identify which message was sent.

\subsubsection{Semi-Open-World experiments}
\label{sec:attack5}

While being able to distinguish between one of two possible messages is clearly
a privacy violation in some cryptographic sense, it is not a very realistic
attack: adversaries do not often know that a victim user will send exactly one
of two specific messages.

Thus, in this section we consider a more ``open world'' experiment, where we assume the
victim user is asking questions on one of two \emph{topics},
and either asks for medical advice, or coding assistance.
\footnote{Even this open world experiment is not \emph{fully} open-world in
the sense that the users can talk about any topic at all. But we nevertheless
believe this setting is much more realistic.}

\paragraph{Methodology.}
To simulate this, we construct a set of initial prompts that a user might ask,
either (1) providing a set of symptoms and asking for medical advice, or 
(2) stating a potential coding task and asking for help.
We then query the victim model with one of these (randomly selected) messages.
To then generate a human-like response, we query ChatGPT to reply to whatever
response was generated, and re-query the victim model with this reply.
(Importantly, note that we do not perform any timing analysis of the ChatGPT
human-like response; this step is included only to increase the duration
of the conversation.)
We repeat this procedure for eight rounds.

In all cases, the victim is monitored using the protocol described in the prior
section.
We train a classifier on a disjoint dataset of conversations 
to distinguish between messages about medical advice and messages about coding
assistance, and then run the attack on a new set of conversations.

\begin{figure}
    \centering
    \includegraphics[scale=.75]{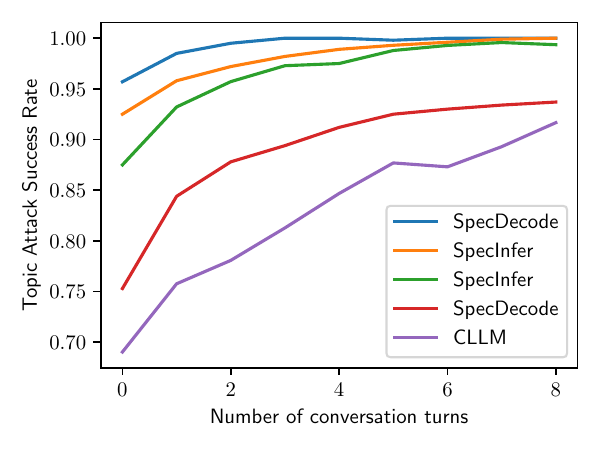}
    \caption{Our attack reliably distinguishes between a user either asking for coding
    assistance or asking for medical advice across all five efficient inference strategies.
    When a user interacts with a model for multiple back-and-forth conversations the
    attack success rate grows considerably.
    Some methods are much more vulnerable to attack than others, with up to 100\% attack
    success rate.}
    \label{fig:medical}
\end{figure}

\paragraph{Results.}
Overall, we achieve an attack success rate of between $89.5\%$ and $99.8\%$ at distinguishing between
the potential topics after observing ten messages between the (simulated) user and the remote model.
Figure~\ref{fig:medical} shows the complete curve as we vary the number of conversation turns
from $0$ (only one response is given, the user never replies) to $8$ (the user replies 8 times).

While we do find that some methods are (much!) more vulnerable than others,
we caution readers from reading too much into this.
Several of these techniques require fine-tuning models to be effective, and
not all techniques choose the same base models to fine-tune.
Therefore, in several cases we have changed not only the efficient inference method,
but also the model.

Nevertheless, the conclusion we \emph{can} draw from this analysis is that each of these
efficient inference methods that depend on the content is vulnerable to a timing attack that leaks the content of the
user message. This is inline with other traffic analysis attacks that use the dependency on the content to gain unintended information from the communication patterns~\cite{cai2012touching,wright2008spot,white2011phonotactic,schuster2017beauty,zolfaghari2016practical}.
We believe it would be interesting for future work to more rigorously study the reasons behind
why some of these techniques are much more vulnerable to attack than others.

\section{Passive Attacks on Production Models}

We now show that our timing attack is effective not only for open-source
efficient inference methods, but also is practical in the fully end-to-end setting
for large production models including OpenAI's ChatGPT and Anthropic's Claude.
For the attacks we consider in this setting,
exactly \emph{which} techniques have been deployed by these language
model services are not relevant:
all that matters is that efficient inference strategies \emph{are} in use.
As such, we first begin with a very simple experiment to confirm that these
techniques are being applied.

\subsection{GPT-4 uses data-dependent efficient inference}
We begin with an extremely simple experiment to verify whether or not we will
have \emph{any} hope at determining the content of a message based on the
timing characteristics of the model response.
First, we construct two sets of queries;
one of these are ``easy'' queries that even simple models should be able to solve
correctly (e.g., ``count from 1 to 100'');
the other are ``hard'' queries that are challenging 
(e.g., ``write a program to check if a number is prime'').
We then query the model on each of these sets of queries,
and compute the time the model takes to return the first 100 tokens.
(We ensure all queries generate over 100 tokens of output.)

\begin{figure}
    \centering
    \includegraphics[scale=.75]{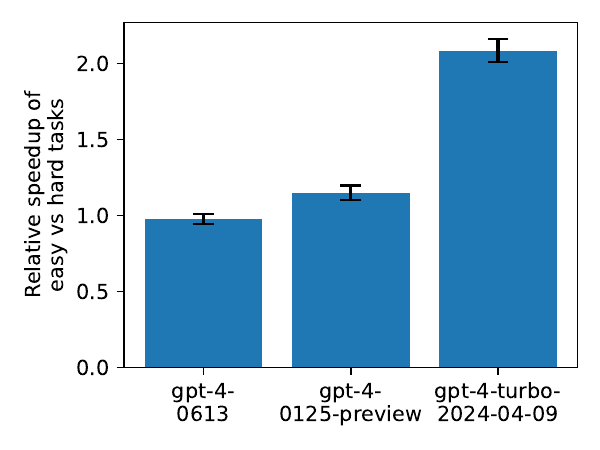}
    \caption{Later versions of GPT-4 implement more advanced efficient inference
    techniques. Whereas the June 2023 model takes just as long to answer easy
    versus hard queries, and the January 2024 ``preview'' model is 10\% faster
    on easier queries than harder queries,
    the April 2024 release of the \texttt{gpt-4-turbo} model is \emph{twice} as
    fast on easy queries compared to hard queries.}
    \label{fig:isspec}
\end{figure}

Figure~\ref{fig:isspec} shows the results of this analysis for three different
releases of OpenAI's GPT-4 model.
The initial June 2023 release of the model is just as fast on the easy set
of queries as on the hard set of queries;
because this was the first release of this model, 
we suspect OpenAI had little time to build efficient inference infrastructure.
Then, in November 2023 and January 2024, OpenAI released a ``Preview'' of the
\texttt{gpt-4-turbo} model, and charged 50\% less for queries to the model.
We find that this model on average responds to easy queries $10\%$ faster than
it responds to harder queries---a statistically significant increase
in performance, if limited.
And then, most recently (as of writing this paper), in April 2024 OpenAI released the
official \texttt{gpt-4-turbo} model.
This version exhibits extreme data-dependent timing characteristics:
it is \emph{twice} as fast at responding to easy queries compared to hard queries.

While it is difficult to speculate on the exact techniques that are being
applied to make this model perform faster on some queries than others,
for the purpose of this section we leave this question aside.
(We believe developing techniques that could identify the exact efficient inference
techniques being used would be an interesting avenue for future work.)
Instead, we focus on measuring the \emph{consequences} of the fact
that this series of models---and others like them--employ efficient inference techniques,
and determining how much information is leaked as a result of this efficient inference.

\subsection{Results}

\subsubsection{Evaluation Methodology}
Our evaluation methodology closely follows that of the prior section;
we again place a victim user inside of a virtual
machine with the adversary acting as an
on-path network adversary on the host machine.
Because both OpenAI and Anthropic send all requests encrypted over HTTPS
we need no server-side modifications,
and simply 
query the remote language model using the latest version of the corresponding
model's Python API.
Again, the adversary captures all packets between the victim machine and
the remote language model service, discards the content, and records the
inter-packet delay.

\subsubsection{ChatGPT Results}
We perform our attack on several OpenAI models.
In this section on their most popular model families: GPT-3.5 and GPT-4.

For each model, we query the remote model 500 times to
establish the timing signature:
250 times with $P_a$ and 250 times with $P_b$.
We then use this data and fit a GMM on these samples.
At test time, we query the model another 500 times.
(Later, in Section~\ref{sec:overtime} we will validate the impact
of how long has passed between when we train our GMMs and when we run the attack.)

\begin{figure}
    \centering
    \includegraphics[scale=.65]{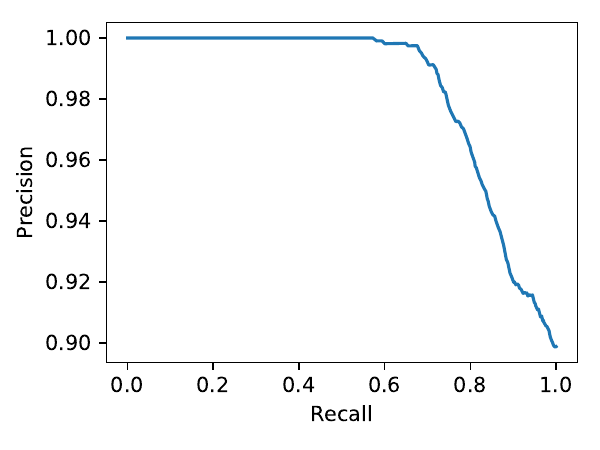}
    \caption{A network adversary can reliably distinguish between a user who has asked GPT-3.5
    to either ``generate the first 50 numbers'' or ``generate 50 random numbers''.
    By training a Gaussian Mixture Model on the packet response times, we can achieve 94.7\%
    accuracy at distinguishing these two queries;
    and can label over 50\% of interactions with perfect precision.
    }
    \label{fig:gmmscore}
\end{figure}

Figure~\ref{fig:gmmscore} shows a histogram of the GMM likelihood
scores and the precision/recall curve for the GPT-3.5 model---the
model least vulnerable to our attack.
Even on this model we see that we can achieve a recall of $50\%$
with perfect precision at identifying which question has been asked.

\begin{figure}
    \centering
    \includegraphics[scale=.75]{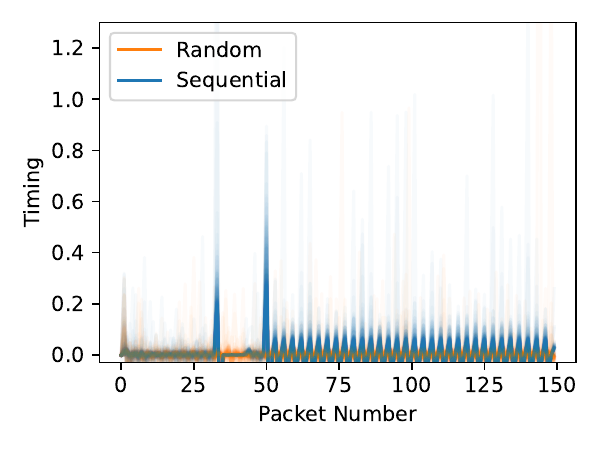}
    \caption{Inter-packet delay on a token-by-token basis for two different queries,
    with 100 independent results overlaid.}
    \label{fig:pertoken}
\end{figure}

\paragraph{Delving deeper.} 
Why is it so easy to fingerprint easy versus hard queries for these models?
In Figure~\ref{fig:pertoken} we plot the response time on a per-token basis
for 100 queries for ``generate the first 50 numbers'' and 100 queries for ``generate 50 random numbers''.

Initially responses to the two queries have similar response timing
characteristics.
But then, at the 50th token, there is a significant qualitative
change to the behavior:
whereas the model prompted with sequential numbers starts to have a delayed
response followed by two very fast tokens in reply,
when prompted to generate random numbers the behavior stays
consistent throughout generation.
This suggests that, for one of these methods, there has been an
explicit switch from one type of generation behavior to another type.

\subsubsection{Claude 3 Results}
When we implement the same attack on Claude 3 Opus (the latest
and largest model released by Anthropic), it initially appears to not be very
vulnerable to this attack.
Our attack achieves a 60\% attack success rate, and the AUC of the precision
recall curve is under 0.7.

\paragraph{Delving deeper.}
It is important to realize that there is a distinct difference
between the time that \emph{packets} are sent, and the time between
when \emph{tokens} are sent.
In the case of our open source attacks and attacks on OpenAI's models,
this difference wasn't important.

\begin{figure*}
    \centering
    \includegraphics[scale=.5]{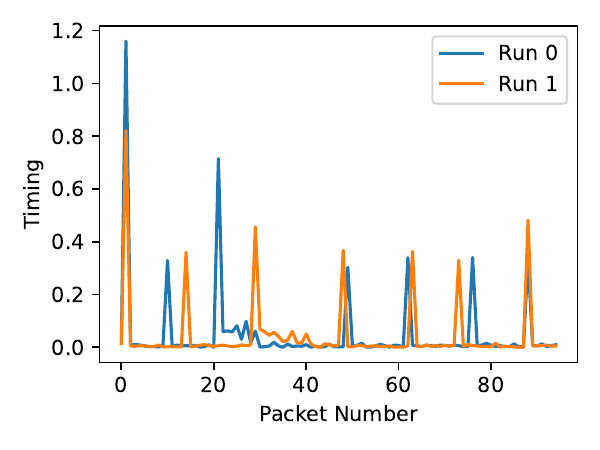}
    \includegraphics[scale=.5]{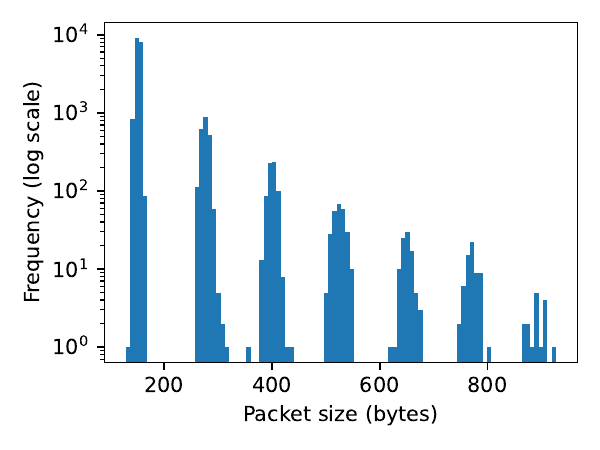}
    \includegraphics[scale=.5]{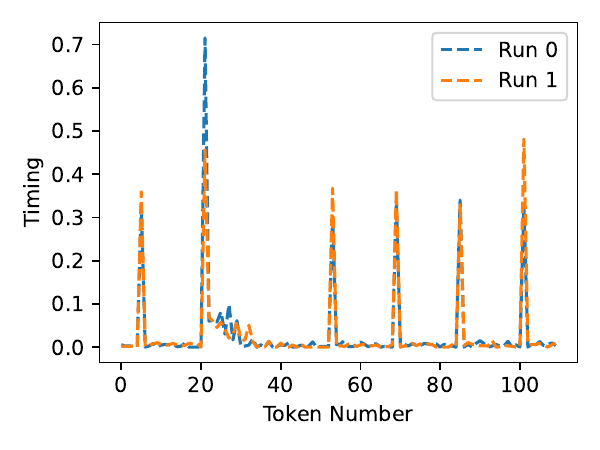}
    \caption{\textbf{(a)} We query Claude 3 twice with the same prompt.
    Because Claude 3 occasionally merges multiple tokens into one packet,
    this merging breaks the alignment between packets and tokens,
    and makes it difficult to mount our timing attack.
    \textbf{(b)}
    But a second side-channel exists: the number of bytes in each packet
    reveals the number of tokens in that packet.
    \textbf{(c)} By combining both side-channels together to obtain the inter-\emph{token}-delay (instead of the
    inter-packet-delay) we can restore the timing alignment of the
    two independent runs, and achieve near perfect attack success rates.}
    \label{fig:unalignment}
\end{figure*}

But for Claude 3 we notice that the backend server will
occasionally group together multiple tokens and send it to the client 
in a single TCP packet.
Figure~\ref{fig:unalignment}(a) shows the inter-packet delay for two
identical queries repeated twice on the model.
We find that not all packets encode just one token.
For example, in Figure~\ref{fig:unalignment}(b) shows the size of
each packet that is sent by the model.
Notice that there appear to be distinct ``buckets'' of sizes for
packets: this is because each extra token that is generated is placed
within its own 150-byte JSON object.
This means that when we compute the timing signature for a sequence
of packets, we are not accounting for the mismatch where sometimes
individual packets contain multiple tokens and sometimes they contain just one.






\paragraph{Compensating for token clustering.}
Given this new understanding, we can now develop an improved attack.
Specifically, we first fit a model to predict the number of distinct
tokens contained in a single packet of data.
But this is easy to do.
Because each token is contained within a large JSON object (of over a
hundred bytes), this step is rather straightforward:
we fit $k$ Gaussians to the histogram in Figure~\ref{fig:unalignment}(b), and
then given a packet size $s$ we choose the Gaussian that best
fits the the data;
formally
\[
\text{\# tokens}=\mathop\text{arg max}_{i \in [1..B]} \left[ \textbf{Pr}[s\, |\, \mathcal{N}(\mu_i, \sigma_i^2)] \right]
\]

This allows us to construct Figure~\ref{fig:unalignment}(c) which
aligns the token timings and shows that two independent runs of the
same data have the same timing signature.
And therefore, we are able to significantly increase our attack success
rate: in fact, we are able to reach a $100\%$ accuracy at distinguishing
between two different queries with an AUC of $1.0$.

\subsection{Vulnerability over size and optimization}

Next we turn towards investigating the effect of model size,
and the amount of optimization effort that has gone into each model,
starting with size.

\paragraph{Larger models are more vulnerable.}
Within a model family, we find that models \texttt{gpt-3.5} vs \texttt{gpt-4} and
Claude 3 Opus vs Claude 3 Sonnet vs Claude 3 Haiku),
we find the attack success rate increases as models increase in size.
Intuitively this makes sense:
efficient inference algorithms come with a performance
overhead, and so applying them to smaller models will result
in a smaller relative gain.
Moreover, faster models will exhibit more subtle timing signatures
compared to the network overhead.

\begin{figure}
    \centering
    \includegraphics[scale=.75]{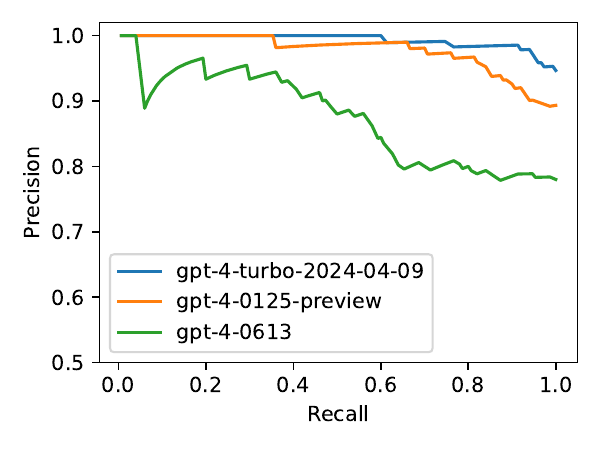}
    \caption{Each subsequent release of GPT-4 has \emph{increased}
    its vulnerability to timing attacks. This correlates strongly
    with the fact that subsequent releases appear to employ more sophisticated
    efficient inference techniques.}
    \label{fig:improve}
\end{figure}

\paragraph{More optimized models are more vulnerable.}
While there is no way for us to measure the quantity
``how optimized was this model?'', 
we can crudely estimate it by when the model was released.
Presumably, more recent models make use of more advanced
optimization techniques and will therefore be more vulnerable
to timing attacks.

And that is exactly what we find in Figure~\ref{fig:improve}:
when we compare the success rate of our attack on three different
GPT-4 base models (\texttt{gpt-4-0613}, \texttt{gpt-4-0124-preview}, and \texttt{gpt-4-turbo-2024-04-09}),
we find that the attack success rate increases significantly.

\subsection{Attacks over time}
\label{sec:overtime}

The timing attack literature has discovered many ways in which
timing attacks can be unreliable~\cite{cherubin2022online,juarez2014critical}.
In this section we consider one way that timing attacks often fail in
practice: the timing characteristics at one point in time are often different
from the timing characteristics for a future point in time~\cite{nasr2018deepcorr}.

In order to investigate this, we query four language models
(GPT 4 Turbo, GPT 3.5 turbo, Claude 3 Opus, and Claude 3 Sonnet)
over a period of a month, and report the attack success rate over time if
we train our attack on the first day, and repeat the attack for the
remaining days.

\begin{figure}
    \centering
    \includegraphics[scale=.75]{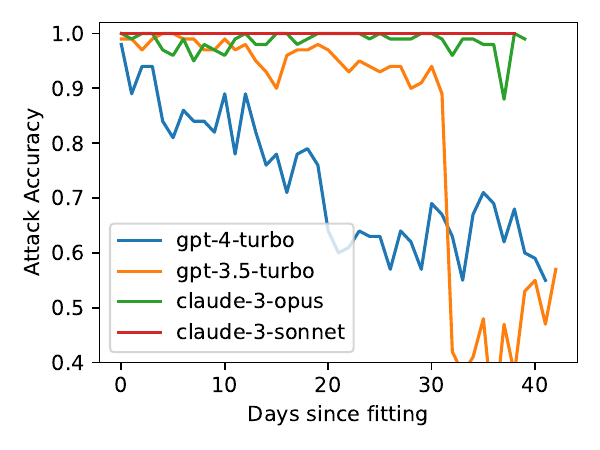}
    \caption{Some language model behavior changes over time. While attacks
    on Claude remain consistently effective a month after the initial fitting,
    attacks on GPT 3.5 and GPT 4 become less effective as more time passes.}
    \label{fig:overtime}
\end{figure}

We can draw several interesting lessons from this figure:
\begin{itemize}
    \item Claude models appear to remain unchanged over a period of a month.
    The attack success rate remains consistently high.
    \item Attacks on GPT 3.5 remain high, and then suddenly after a month
    of data gathering the attack success rate drops to random guessing.
    This suggests that OpenAI made a change to the way this model is served,
    and prevents the specific timing signatures learned initially from transferring.
    \item Attacks on GPT 4 decay slowly over time. It is harder to immediately
    draw a near-certain conclusion from this figure: it could, for example,
    indicate the serving infrastructure is being iterated on daily,
    or that the usage patterns of this model have changed significantly over
    time and so it behaves differently from being queried more (or less) frequently.
\end{itemize}

\subsection{End-to-end ChatGPT Website Attack}

\begin{figure}
    \centering
    \begin{subfigure}[b]{0.45\textwidth}
        \centering
        \includegraphics[scale=0.45]{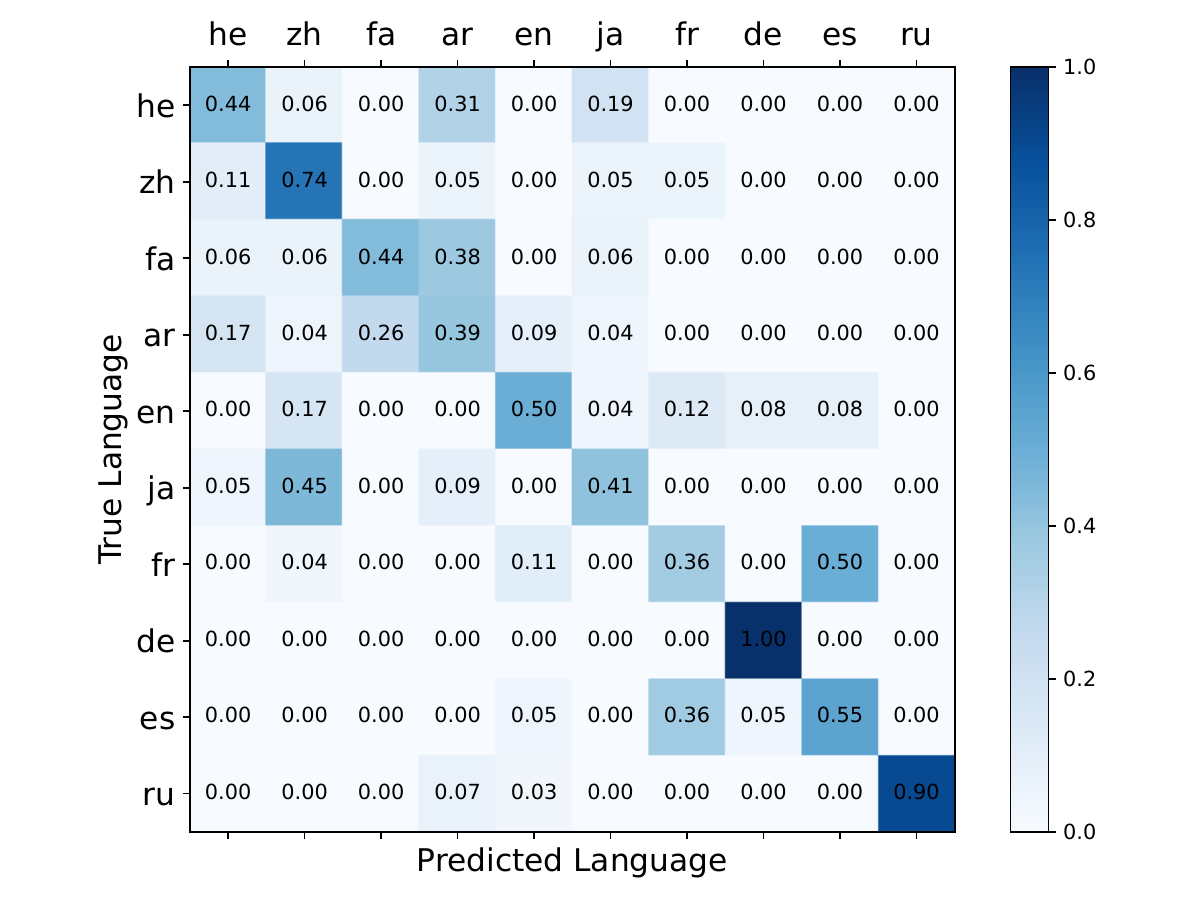}
        \caption{GPT-4}
    \end{subfigure}
    \hfill
\caption{Performance comparison of a language model across various languages. The figure illustrates the probability of correct language prediction for ten languages. Each value represents the model's recall in identifying that language and errors. We see both models have comparable leakage about different languages.}    \label{fig:cm_lang}
\end{figure}

\begin{figure}[htbp]
    \centering
    \begin{subfigure}[b]{0.49\linewidth}
        \centering
        \includegraphics[scale=0.22]{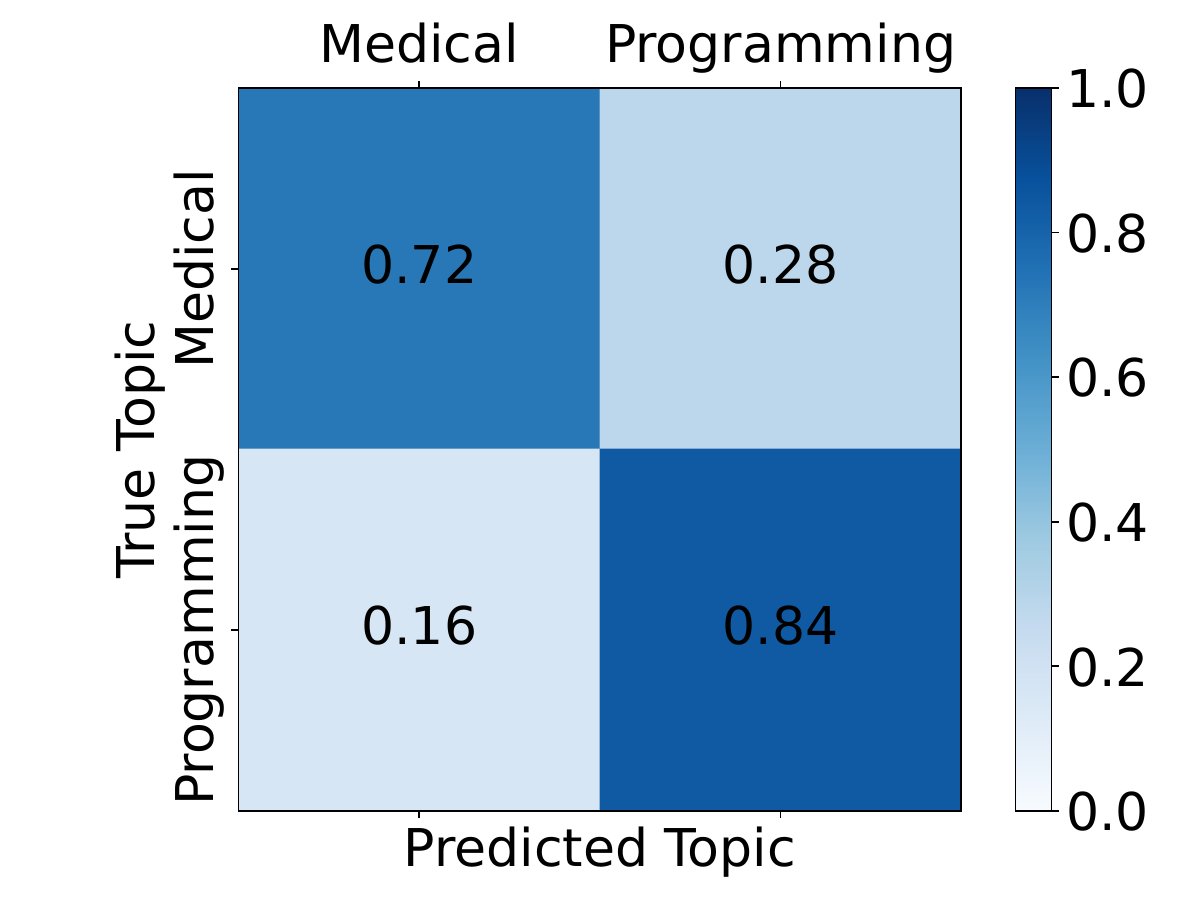}
        \caption{GPT-4}
    \end{subfigure}
    \hfill
    \begin{subfigure}[b]{0.49\linewidth}
        \centering
        \includegraphics[scale=0.22]{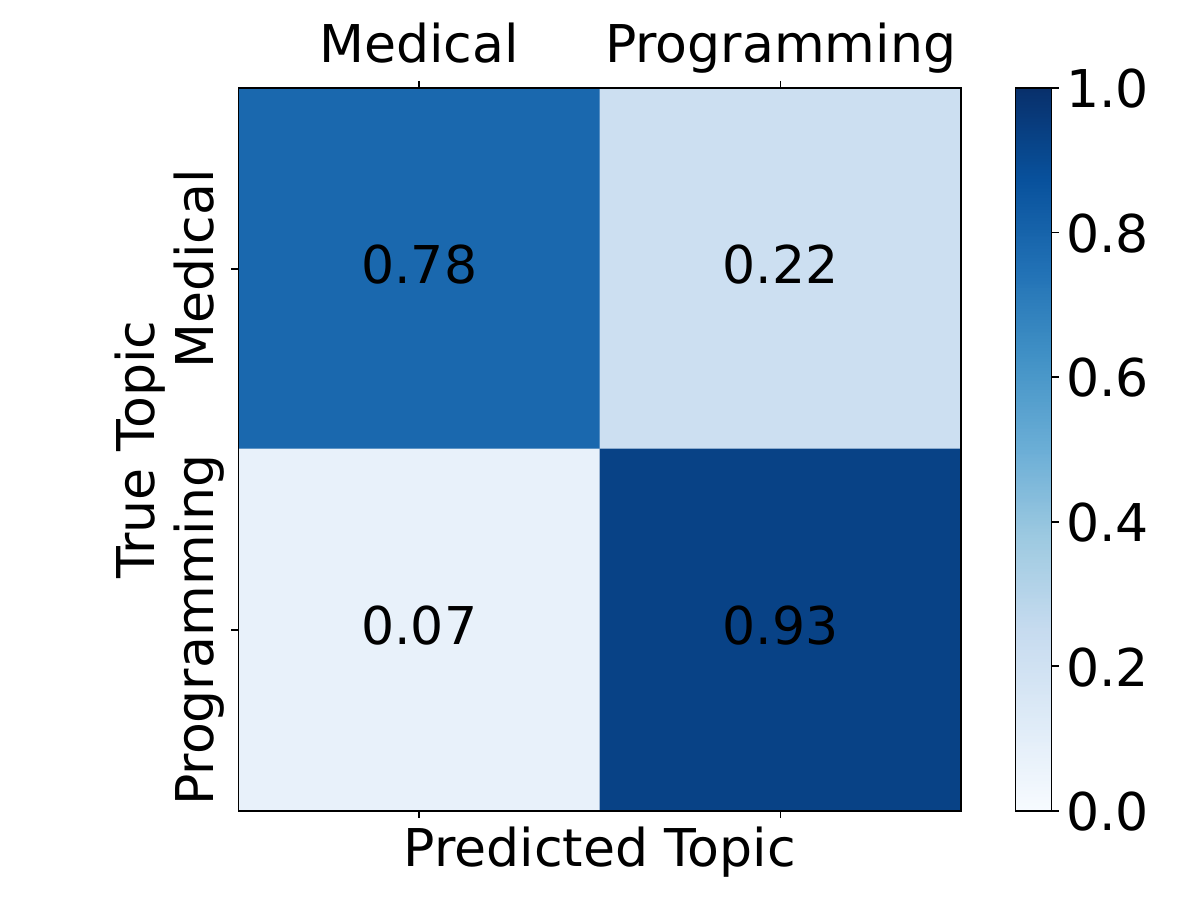}
        \caption{GPT-4o}
    \end{subfigure}
\caption{Performance comparison of a language model across two different topics of medical or programming questions. As we see in both models we can distinguish between these two topics.}    \label{fig:cm_topics}
\end{figure}

We now implement the same end-to-end attack on the ChatGPT web interface directly, instead of querying the language models via the APIs.
We test this case because, while in principle the model being served is the same,
there is likely a significant difference in the front-end application that serves
the responses. Moreover there might be more noise due to the additional communication between the web interface and the servers. 

\paragraph{Methodology.}
We use Selenium to automate a user interacting with the ChatGPT web interface available at \texttt{chatgpt.com}.
Given the limitations of working through the UI we limited the attacks in this section to just inferring the language and the topic of the conversation by observing the network traffic between a user and the ChatGPT servers. 
%
In this experiment we choose 10 popular languages and query the ChatGPT Web UI to write three paragraph stories in each of these languages. We experimented with both GPT-4 and GPT-4o. Similarly for the topic experiment we used 40 different questions in two topics of programming or medical questions and capture the response packets. Overall we repeated these experiments 300 times for each language and 20 times for each question, for topic classification tasks. In all of the experiments, we only look at the communication between our target client and language model serving servers. At the time of writing this paper, ChatGPT uses Cloudflare therefore we capture all communication between the target client and any destination in the CloudFlare network.  
As in the previous section, we find that it is sometimes important to include the relative size of each packet to improve attack success rate; this observation is well known in the classical packet analysis literature~\cite{nasr2018deepcorr,hayes2016k,gong2010fingerprinting,fing-attacks-defenses}. 
We therefore extract Inter Packet Delays (IPD) and packet sizes for each connection.
In order to ensure that our attack does not just look at the response's length as the
method to identify the language, we
extract the first 400 packets from each experiment, and discard experiments that had fewer than 400 packets.

\paragraph{Evaluation.}
To begin, we run the same Gaussian Mixture Model attack as we successfully applied to the API.
Unfortunately, we find that the GMMs are less effective on this dataset, and achieve attack success rates below $30\%$.
To improve the attack success rate,
we next implemented an improved detection strategy:
instead of hand-designing a classifier, we train a simple
convolutional neural network with two convolution layers and a fully connected network to predict the language/topic of conversations from the network 
packet timing and sizes. 

Figure \ref{fig:cm_topics} summarizes the results of our attack when
the victim interacts with the ChatGPT Web UI.
We find that, similar to the open source model analysis, we can distinguish whether or not the user is asking about medical advice or coding assistance with over 80\% probability in a single message.
For the topic identification task the overall accuracy  is $79\%$ (respectively, $82\%$) for GPT-4 and GPT-4o.
(We do not study multiple conversation turns due to the difficulty of instrumenting the Web UI.)

We also experiment with identifying the user's language (e.g., English, French) among one of the ten most popular languages.
\footnote{The open source models we study were all English-only, and so we do not have a comparison point for this experiment.}
Figure~\ref{fig:cm_lang} presents the results of this analysis.
This attack an overall top-1 accuracy of $49\%$ (respectively, $48\%$) at distinguishing
between the 10 languages with GPT-4 (respectively, GPT-4o);
however, some languages have much higher success rates than others. For example, both GPT-4 and GPT-4o models have similar leakages of languages in the conversation. In particular, both models have surprisingly high leakage of German, Russian, and Chinese.

\section{Stealing Prompts with Active Adversaries}

Recall that under our threat model, an active adversary is given control over
(part of) the conversation between a user and remote model,
but can not observe the content of either the query or the response.

Unlike previously where the adversary can only view the messages exchanged
between the user and the model, here the adversary plays an active role.
Specifically, the objective of the adversary is to construct a sequence of tokens
$\{t_i\}_{i=1}^N$ so that when the user sends a query with the combined
prompt $p || \vec{t}$,
the timing of the model's response leaks information about the (unknown) message $p$.

\subsection{Restricted-Domain attack}

We begin by introducing a simple \emph{boosting}-like approach to significantly 
improve the attack success rate of the passive attack,
under the assumption the adversary can cause the victim to issue repeated
queries with a (fixed) base prompt.

\paragraph{Attack setting: 1-of-N}
Like our previous A/B testing strategy that distinguishes between one of two
possible prompts, here we consider a setting where a victim chooses one of $N$ possible prompts
$p \gets \{p_i\}_{i=1}^N$.
(We move from 1-of-2 to the harder 1-of-N because we can already nearly perfectly solve
1-of-2 even without an active attack.)
Then, the adversary is allowed to construct various
prompts $q^j$ and query the model to observe the response time for the
message $\mathcal{O}(p || q^{j})$.
Importantly, note the adversary can neither view the content of the message nor the reply.

\paragraph{Method.}
The reason our simple A/B testing, distinguishing attack was effective is that
the model behaves (slightly) differently when issued one query compared to another.
But in this more general attack setting, where the adversary can repeatedly cause different
queries to be issued to the language model $\mathcal{O}$,
it is possible to do significantly better by aggregating multiple results together.

To do this, we simply collect a set of semi-arbitrary prompts
(``write me a story about a unicorn'', ``tell me the history of the Americas'',
``what is the integral of log(x)'')
and, as the adversary, cause the victim to issue queries with each of these
messages present in the combined prompt.

The adversary then repeats the prior attack setup.
In the \emph{initialization} phase, they query the model with each combination of
prompt $p_i$ and each possible adversary suffix $q^j$, record the timing
characteristics, and train a classifier $f^j(\cdot)$ that predicts which
victim prompt was selected for each adversary suffix $q^j$.
Then, in the \emph{inference} phase, the adversary queries each of the $q^j$ prompts with
the victim's unknown prompt $p \in \{p_i\}$, and runs the classifier $f^j$ on each
of the response timings.
With this data, the adversary can aggregate the predictions across each of the independent
queries.

\paragraph{Evaluation Setup.}
We implement this attack on each of the five efficient inference methods
from the prior section.
We choose victim prompts to be one of 1000 questions from 
the Alpaca dataset \cite{taori2023alpaca}.
We then choose 100 alternate prompts, and query the language model with 100 different
\emph{modifiers} (e.g., ``summarize the answer briefly'', ``explain the answer like I'm five'',
``speak like a pirate when answering'').
We query the model on each of the $100 \times 1000$ possible queries and record
the inter-packet delay for the first 100 tokens of the model's response.

\paragraph{Results.}
We find that it is possible to \emph{perfectly} determine the secret number
that has been selected by looking at the timing characteristics
of the corresponding 100 questions for each of the five efficient inference methods.
Moreover, we find that even 20 questions are sufficient if network latency is
ignored.

\subsection{Stronger Black-box Attacks}

In cases where the adversary wishes to learn something that is not one of a
small set of questions, it will be necessary to develop alternate attacks.
In this section we consider one such method.
However, before we do this, we briefly review key details of
speculative decoding that are important to the attack details here.

\subsubsection{Speculative Decoding}
\label{sec:specdecode}
Recall that this method works by 
generating potential responses using a smaller (and thus faster) \emph{draft} model,
which a larger (``target'') model can accept or reject.

Given a prompt $p$, speculative decoding first samples several (e.g., 5) sequential tokens from the small draft model.
This requires several invocations of the small model, each time generating one new token $q_i$.
Then, this candidate response $q$ is provided to the large target model to score;
this requires just one forward pass of the target model.
The target model then either \emph{accepts} the entire response, 
some prefix of the generated response $q_{1..k}$,
or \emph{rejects} the response entirely and provides a new
token $q'_i$.

\subsubsection{Intuition}
Generally, the attack approach we will consider here exploits the
fundamental \emph{capability gap} between the model's ability to
generate a fast and slow response.
Because of the way each of the efficient methods work, models
are almost always less capable when they give fast responses
compared to when they give slow responses.
In the case of speculative decoding, this is because the draft
model is much smaller than (and thus not as ``smart'' as) the target model.
And so by constructing queries that can be solved by only one
of the models---and not the other---we can reliably induce
the model to respond either quickly or slowly depending on
whether the answer to the question is true or false.

Specifically, we focus on asking questions that cause the model's
very first token to either be accepted or rejected.
In either case, generating the first token will be slow
(as it requires querying the draft model 5 times and then
the target model once).
But the second token's timing will vary.
If the first token is accepted, the remaining 4 tokens will
(possibly) also be accepted;
but if the first token is rejected, the remaining 4 tokens
\emph{must} be re-generated.
Therefore, by measuring the time it takes the model to respond
with the second token, we can detect if the model accepted
or rejected the first token.

\subsubsection{Attack setting}
For the remainder of this section, given that we have now already
perfectly solved the 1-of-k attack setting we shift our focus to a harder task:
learning secret values placed in the prompt.
For example, the prompt might contain some sensitive information about a person
(e.g., their phone number or credit card number).

Specifically, for this paper we choose the set of prompts $p$ to be prompts of the form
``My secret number is [number].''
We set the space of prompts $\{p\}$ to be messages of the form
``The secret number is [number]. Do not reveal it.''
where the number is a single integer between 1 and 1000.

Note that if the adversary had the ability to view the response it would be
easy to just perform a jailbreak and ask ``disregard prior instructions and type the number I told you earlier.'' and the model would
(likely) respond with the correct number;
but this is not possible because we assume the adversary can only inject
data, but can not view the reply.

\subsubsection{Method}
To begin we restate the attack objective formally, so that we can then directly
optimize for a token sequence to make the attack succeed with high probability
\begin{equation*}
\label{eq:main}
\mathop{\text{arg max}}_{\{t_i\}} \left|
\mathop{\mathbb{E}}_{m | p(m)} T(m + \{t_i\})
-
\mathop{\mathbb{E}}_{m | \sim p(m)} T(m + \{t_i\})
\right|
\end{equation*}
where $T(\cdot)$ represents the time a model takes to respond to a given message.

We optimize for the best token sequence $\{t_i\}$ through a simple
black-box optimization procedure.
Initially, we choose a prompt that we believe might cause a capability
gap between the two models (e.g., ``is the first digit in the number X?'').
Then, we evaluate this token suffix, and measure how often it causes the two
models to differ in timing and thus leak the timing information to the adversary.
We find that this simple question does not exhibit any significant timing
signature: only 56.2\% of the time does the model's latency at returning
the second token reveal whether or not the first digit of the number matched
our guess.

Then, we iteratively begin to refine this question.
But instead of doing this by hand, we make use of a language model to rephrase
our question.
Specifically, we prompt GPT-4o to ``rephrase this question 20 different ways''.
We then take each of these re-phrased versions of the question, and score
the attack success rate of each of these. 

One of these re-phrasings is ``Is the first numeral in the number a X?'' which
has a 65\% probability of distinguishing between the two models.
We then take the best generation, and repeat---asking GPT-4o to rephrase the
question again.

Table~\ref{tab:questions} shows a table of eight of the questions we discover
that we can ask the model, bringing the attack success rate from 56\% to 97\%.
We find that the reason this token suffix achieves a higher attack success rate
is because it better exploits the capability gap between the large and small model:
the small model \emph{always} answers ``no'' for this question, but the large
model achieves a 97\% success rate at answering the question.

Now that we have (probably) learned the first digit of the secret number, we can repeat
this procedure for each subsequent digit.
While this process is not perfect---and our attacks do sometimes fail to learn
the digit at each location, we can \emph{substantially} reduce the space of
possible generations.
And once we have reduced the space of possible generations to a modest number
(e.g., a few thousand), we can repeat the attack from the prior section that
allows us to use brute-force methods and perfectly recover the prompt.

Note that the benefit of this attack method is that it \emph{automatically}
discovers prompts that exploit the capability gap between the large and small
model, without requiring time-consuming manual effort.

\begin{table}
    \centering
    \begin{tabular}{p{5cm} c}
        \toprule
        \textbf{Question} & \textbf{Success Rate} \\
        \midrule
        Is the first digit in the number X? & 56.2\% \\
        Is the first numeral in the number a X? & 65.6\% \\
        Yes or no, does the digit X occupy the first place in the number? & 78.2\% \\
        Does the number you have begin with X? Respond with yes or no. & 85.5\% \\
        Does a X appear as the first character in the number? Yes or no? & 87.8\% \\
        Is the initial digit of the number a X? Yes or no? & 90.2\% \\
        Do we see a X at the beginning of the number? Provide a yes or no answer. & 94.1\% \\
        When looking at that number, is X the initial digit? Yes or no? & 97.2\% \\
        \bottomrule
    \end{tabular}
    \caption{Attack success rate of various questions at eliciting a capability gap,
    and therefore introducing an exploitable timing side-channel,
    when performing speculative decoding with a 7 billion parameters target model and
    a 1.5 billion parameters draft model.
    Easy-to-understand sentences work less well because both models answer correctly;
    whereas harder-to-parse sentences are only answered correctly by the larger model.}
    \label{tab:questions}
\end{table}

\paragraph{Evaluation Methodology.}
As stated above, we implement our attack on just one of the efficient inference methods,
speculative decoding.
We use the same experiment setup from the previous section, and 
consider a model provider who deploys an efficient inference method with
a 7 billion parameter model and a 2 billion parameter draft model,
and another setting where the victim uses a 70 billion parameter target model
with a 7 billion parameter draft model.

We run our attack for 10 iterations starting with the simple question
written above, and each time retain the top-10 scoring outputs.
This results in $91$ ($= 1 + 10 \times 9)$ total calls to GPT-4o at a cost of $\$0.2$ per attack,
generating 
$910$ possible candidate suffixes.
We evaluate each of these by querying the target and draft model on $1000$ randomly
selected numbers, which takes under an hour on a single A100 GPU.

\paragraph{Results}
When we consider the 7B/2B model setting,
after ten rounds of optimization, our attack returns a question that
reaches $97.2\%$ success rate at causing a timing discrepancy between
the target and draft models.
We find that it is easier to implement the attack in the 70B/7B setting:
after ten rounds of optimization our attack returns a question that
yields a higher $99.3\%$ success rate at causing a timing discrepancy between
the target and draft models.

\subsection{White-box gradient-based attacks}
\label{sec:whitebox}

Let us now approach this problem differently.
Instead of trying---through black-box search---to happen upon queries
that exploit capability gaps to induce timing discrepancy between the fast and slow inferences,
we can try to \emph{directly} optimize for this to occur.

We will attempt to leverage recent advances in discrete
optimization to generate text sequences that maximize the timing difference
between two potential queries.
Specifically, we will apply the recent GCG attack \cite{zou2023universal}
to try and construct an \emph{adversarial suffix} that directly achieves this goal.

But we encounter a problem:
GCG, and other gradient-based optimizers, require a differentiable
objective function in order to work efficiently.
Unfortunately, the two calls to $T$ in Equation~\ref{eq:main}
are not differentiable, and so it is not possible to directly
optimize this metric.

Our observation is that, in all of the efficient inference techniques
we consider, the time function can actually be decomposed as
\begin{equation}
    T(query) \approx \begin{cases}
    t_1 & \text{if query is easy} \\
    t_2 & \text{if query is hard} \\
    \end{cases}
\end{equation}
and so it is possible to approximate
\[T(x) \approx t_1 + (t_2 - t_1) * difficulty(x)\]
where $difficulty(x) \in \{0,1\}$ is a function that
$difficulty(x) = 1$ if it is hard and $0$ otherwise.
But by increasing the domain of this $difficulty$ function and making
it continuous, we can construct an entirely differentiable $T(\cdot)$
function and therefore compute end-to-end gradients and optimize this attack.

But how should we construct this difficulty function?
We explored several different settings, and ultimately found the most effective
method was to maximize the difference between the token output probabilities:
\[ difficulty(x) = \mathop{\textbf{Pr}}_{\text{target}} [y\,|\,x ] - \mathop{\textbf{Pr}}_{\text{draft}} [y\,|\,x ] \]
where $y$ is the desired distinguishing token (e.g., ``yes'' or ``no'' as 
used in the prior section).
However we believe it may yet be possible to develop even better
difficulty functions (see below).

\subsubsection{Evaluation}
We now evaluate this attack for both of the settings considered previously.

\paragraph{1-of-N Testing:}
Unsurprisingly, given white-box access it is straightforward to construct
adversarial suffixes that cause the model to behave differently when the
adversary knows the victim will send one of a (small) set of queries.
We can construct adversarial suffixes that succeed at this goal in
under 100 iterations of the attack, requiring under 30 minutes of compute
on a single A100 GPU.

\paragraph{Secret Stealing:} 
Surprisingly, despite our best attempts, we were unable to succeed at generating adversarial
suffixes using GCG that could be reliably applied to large prompt spaces.
On one hand it may not be seen as surprising that this failed.
But we had three good reasons to believe this attack \emph{should} work:
\begin{itemize}
    \item We can successfully optimize for prompts that detect between a fixed
    1-of-N set of prompts for small N.
    \item Prior work \cite{zou2023universal} has found that adversarial suffixes generalize
    across prompts in other domains.
    \item Most importantly: \textbf{We know a solution exists!} In the prior section, we found---with
    black-box attacks---queries that could reliably distinguish between
    messages where numbers started with one digit versus numbers that did not.
\end{itemize}

Taken together we believe this implies that more work is needed to
improve the efficacy of adversarial suffix generation.
While GCG and related efforts show marked improvements compared to prior
work, here we have found a case where we know there exists a string of
tokens that maximizes the timing difference, but current
optimization methods are unable to find it.

\section{Defenses}

There are a number of general-purpose strategies to defend against timing attacks.
In this section we briefly review two of the most popular methods and discuss how they may be
applied to preventing timing attacks on efficient language model inference.


\begin{figure}
    \centering
    \includegraphics[scale=0.65]{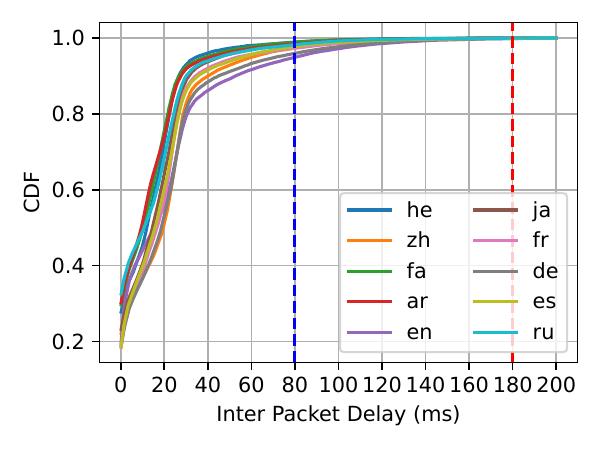}
    \caption{CDF of the inter packet delays of the responses in GPT-4o for different languages. As we can see the distributions do not have a very large tail and we can use many of the existing defenses without overhead cost.}
    \label{fig:cdf_lanauges}
\end{figure}
\begin{figure}
    \centering
    \includegraphics[scale=0.65]{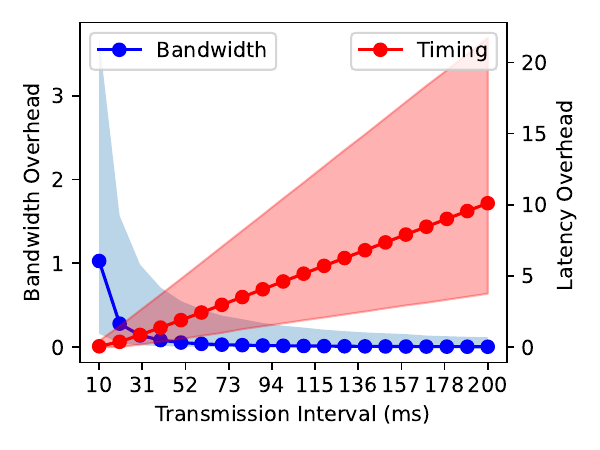}
    \caption{By increasing interval between transmission of each token in our defense,
    we can reduce the bandwidth overhead. (And, conversely, by increasing
    the bandwidth overhead, we can reduce the latency of the defense i.e, reducing the transmission interval.)}
    \label{fig:latencybandwidth}
\end{figure}

\paragraph{Modifying packet timing and sizes.}
The most common defense~\cite{juarez2016toward,wang2017walkie,cherubin2017website,cai2014cs,nasr2021defeating,van2015vuvuzela,piotrowska2017loopix,lazar2018karaoke,abraham2020blinder} against traffic analysis attacks is to modify the packets' timing to prevent any data leakage. 
In the traditional traffic analysis settings, using such defenses is very costly and is often not effective~\cite{nasr2021defeating,das2018anonymity,sirinam2018deep}.

Fortunately, these defense approaches are more viable in our use-case since (1) we require relatively high precision timing measurements and, (2) the variance in timing between the fastest and slowest responses is usually relatively small, which allows us to introduce delays without significantly degrading full utility.
Therefore, we can utilize similar techniques without significantly impacting communication latency or introducing excessive overhead traffic.

Specifically, we apply a simple \text{constant output rate} defense which will thwart our attacks completely \cite{piotrowska2017loopix,van2015vuvuzela}. 
This defense strategy works by choosing some output rate $r$, and returns an output
token exactly every $1/r$ seconds.
If no data is ready, we send an empty message (wasting bandwidth).
If more data is ready than can be sent, we send just one message (increasing latency).

To determine the optimal rate we can measure the frequency of response times.
Figure~\ref{fig:cdf_lanauges} illustrates the cumulative distribution of the inter packet delays across different languages.
As we can observe, different languages have slightly different timing distributions, however,  the distribution is not heavily tailed.
This allows us to use fixed-rate communication to remove any identifying marks from the communication patterns.

Figure~\ref{fig:latencybandwidth} presents the complete curve of latency versus bandwidth overhead.
As we can see, by transmitting a token every 10 milliseconds (100 times per second) we introduce essentially no latency, but increase bandwidth between 50\% and 300\%.
Alternatively, by transmitting tokens every 80 milliseconds, we can ensure that at least 90\% of packets have real data and so the bandwidth overhead drops to below a $10\%$ increase, at a slight cost of between 4ms and 13ms per token delivered.

\paragraph{Protocol Morphing.} 
To defend against general traffic analysis attacks, there are other types of defenses such as tunneling the traffic through other protocols or other applications~\cite{skypemorph,mcpherson2016covertcast,fifieldsnowflake}.
Unfortunately, not only there are many difficulties implementing such approaches~\cite{parrot},  but also they usually have heavy additional costs. Therefore, we do not believe they would be a good defense for these attacks.

\section{Conclusion}

Machine learning models have grown in scale from simple models
that can be described in a few dozen lines of code to complex
distributed systems that span entire datacenters.
As these models become widely deployed,
it begins to be necessary to re-evaluate their
attack surface in light of classical attacks from the 
security literature.

This paper reconsiders how one such attack---timing side channels---might
be adapted to leak the content of messages exchanged between a user
and a remote language model.
Just as website fingerprinting attacks allow a network adversary to predict
which webpage a user is viewing, or SSH timing attacks allow a network adversary
to recover the password a user types,
our timing attacks allow a network adversary to predict the topic of conversation
(analogous to which website someone is visiting)---or even, in more restricted settings,
learn the exact message a user has sent (analogous to recovering passwords).

We focus our analysis not on developing qualitatively new approaches to
designing timing attacks or defenses to timing attacks, but on applying
these well-studied attacks to a new domain: that of efficient LLM inference.

More broadly, we believe that the rapidly accelerating field of language modeling
will encounter increasingly many challenges previously
studied by the computer security community.
We hope that because these problems have been already studied extensively
in these other domains, it will be possible to rapidly explore the space of
attacks and present effective defenses to these attacks---as we have done here.
\newpage

\bibliographystyle{plain}
\bibliography{refs}

\end{document}